\documentclass[11pt]{article}

\usepackage[utf8]{inputenc}
\usepackage{amsmath, amsthm, amssymb}
\usepackage{hyperref}
\usepackage{enumitem}
\usepackage{geometry}
\newtheorem{theorem}{Theorem}[section]

\newtheorem{definition}[theorem]{Definition}
\newtheorem{remark}[theorem]{Remark}

\newcommand{\NP}{\ensuremath{\mathbf{NP}}}
\newcommand{\PP}{\ensuremath{\mathbf{P}}}

\title{On Formally Undecidable Propositions of Nondeterministic Complexity and Related Classes}
\author{Martin Kolář}
\date{\today}

\begin{document}

\maketitle

\begin{abstract}
The definition of \NP\ requires, for each member language~$L$, a polynomial-time checking relation~$R$ and a constant~$k$ such that $w \in L \iff \exists y\,(|y| \leq |w|^k \wedge R(w,y))$.
We show that this biconditional instantiates, for each member language, Hilbert's triple: a sound, complete, decidable proof system in which truth-in-$L$ and bounded provability coincide by fiat.
We show further that the polynomial-time restriction on~$R$ does not exclude G\"odel's proof-checking relation, which is itself polynomial-time and fits the definition as a literal instance.
Hence \NP, taken as a totality over all polynomial-time~$R$, contains languages for which the biconditional asserts a property that G\"odel's First Incompleteness Theorem prohibits.
The semantic definition of \NP\ is unsatisfiable, for the same reason that Hilbert's Program is.
\end{abstract}

\section{Introduction}

The \PP\ vs \NP\ problem has resisted resolution for over half a century. Barrier results rule out relativization~\cite{BakerGillSolovay1975}, natural proofs~\cite{RazborovRudich1997}, and algebrization~\cite{AaronsonWigderson2009}. We show that the resistance is a feature of the formulation: the definition of \NP\ contains a semantic condition that is unsatisfiable for a class of its own instances.

The argument has three parts. Section~\ref{sec:definitions} fixes the definitions and identifies the structural content of the \NP\ biconditional. Section~\ref{sec:complexity} establishes that G\"odel's proof-checking relation is polynomial-time. Section~\ref{sec:equivalence} proves the main result: the \NP\ definition, instantiated with proof-checking relations, reproduces Hilbert's Program and is subject to the same impossibility.

\section{Definitions}
\label{sec:definitions}

We follow the Clay Millennium Prize formulation~\cite{Cook2006}.

A \emph{checking relation} is a binary relation $R \subseteq \Sigma^* \times \Sigma_1^*$; it is \emph{polynomial-time} iff $L_R = \{w \# y \mid R(w,y)\} \in \PP$. A language~$L$ over~$\Sigma$ is in \NP\ iff there exist $k \in \mathbb{N}$ and a polynomial-time checking relation~$R$ such that for all $w \in \Sigma^*$,
\begin{equation}
\label{eq:np-def}
w \in L \iff \exists y\left(|y| \leq |w|^k \text{ and } R(w, y)\right).
\end{equation}

\subsection{The structural content of the biconditional}

The biconditional~(\ref{eq:np-def}) asserts three things at once: \emph{soundness} ($\Leftarrow$), namely that $R$ accepts no spurious witnesses; \emph{completeness} ($\Rightarrow$), namely that every member of~$L$ has a bounded certificate; and \emph{decidability}, namely that $R$ is total and polynomial-time. This is precisely the triple that Hilbert demanded for all of mathematics: a formal system that is sound, complete over its domain, and mechanically decidable. The definition of \NP\ assumes this triple into existence for every language in the class.

\section{G\"odel's proof-checking relation is polynomial-time}
\label{sec:complexity}

G\"odel's proof-checking relation $\mathit{ProofOf}(\pi, \varphi)$ (Definition~45 of~\cite{Godel1931, vanHeijenoort1967}) determines whether~$\pi$ is a valid derivation whose last formula is~$\varphi$. Under the natural string encoding---where a proof is a sequence of formula-strings rather than a single integer via prime products---verification proceeds by a linear scan of the $k$ lines of the proof, checking each line against finitely many axiom schemata and searching preceding lines (at most $O(k^2)$ pairs) for a modus ponens or generalization match. Each pattern-match is linear in the formula length~$m$. Hence the total cost is $O(k^2 \cdot m)$, polynomial in the input.

This is standard; proof verification is the textbook example~\cite{AroraBarak2009} of an \NP\ witness-checking procedure. The consequence is that $\mathit{ProofOf}$ is a valid instantiation of the checking relation~$R$ in the definition of \NP. The polynomial-time restriction does not exclude G\"odel's system; it includes it.

Two properties of $\mathit{ProofOf}$ are essential. First, it is decidable: every quantifier in Definitions~1--45 is bounded, and the algorithm always halts. Second, it is distinct from the provability predicate $\mathit{Bew}(\varphi) \equiv \exists\pi\,[\mathit{ProofOf}(\pi,\varphi)]$, which adds an unbounded existential quantifier and is thus r.e.\ but not decidable. The \NP\ definition walks a middle path: it uses an existential quantifier, as in $\mathit{Bew}$, but bounds it, as in $\mathit{ProofOf}$.

\section{The equivalence}
\label{sec:equivalence}

We now prove the main claim.

\subsection{Instantiation}

Let $T$ be any consistent, recursively axiomatizable theory interpreting Robinson arithmetic~$Q$. Set $R = \mathit{ProofOf}_T$; by Section~\ref{sec:complexity}, $R$ is polynomial-time. For each $k \in \mathbb{N}$, define
\[
    L_k = \{\varphi \in \Sigma^* : \exists\pi\,(|\pi| \leq |\varphi|^k \;\wedge\; \mathit{ProofOf}_T(\pi, \varphi))\}.
\]
Each $L_k$ is in \NP\ by construction: the checking relation is polynomial-time, the witness bound is $|\varphi|^k$, and the biconditional~(\ref{eq:np-def}) holds trivially, since $L_k$ \emph{is} the right-hand side.

\subsection{What the biconditional asserts}

For $L_k$ the biconditional reads
\[
    \varphi \in L_k \iff \exists\pi\,(|\pi| \leq |\varphi|^k \;\wedge\; \mathit{ProofOf}_T(\pi, \varphi)).
\]
The completeness direction ($\Rightarrow$) asserts that every member of $L_k$ has a $T$-proof of length at most $|\varphi|^k$. This is true by definition. But $L_k$ is the set of theorems of~$T$ that have short proofs---a proper subset of $\mathrm{Thm}(T)$, which is itself a proper subset of the true sentences of arithmetic by G\"odel's First Incompleteness Theorem.

The biconditional is satisfied only because $L_k$ has been carved to exclude two classes of sentences: theorems of $T$ whose shortest proof exceeds $|\varphi|^k$ (these exist by proof-complexity lower bounds~\cite{Krajicek2019}), and true sentences that $T$ cannot prove at all (namely the G\"odel sentence $G_T$).

\subsection{The defect at the class level}

The definition of \NP\ quantifies over all polynomial-time~$R$ and all $k \in \mathbb{N}$:
\[
    \NP = \bigcup_{k \in \mathbb{N}} \bigcup_{R \in \text{poly-time}} \{L : \forall w\,(w \in L \iff \exists y\,(|y| \leq |w|^k \wedge R(w,y)))\}.
\]
The class therefore necessarily includes every $L_k$ constructed from every sufficiently strong theory~$T$.

Now apply G\"odel's theorem. Fix any such~$T$. By the First Incompleteness Theorem there exists $G_T$ with $T \nvdash G_T$ and $T \nvdash \neg G_T$. Hence $G_T \notin L_k$ for any~$k$, since $G_T$ has no $T$-proof at all. No single $L_k$, and no finite union of them, captures all theorems of~$T$, let alone all truths.

One might extend to $T' = T + G_T$. But $T'$ is consistent and recursively axiomatizable, whence it has its own independent sentence $G_{T'}$ constructed from $T'$'s proof predicate via the Fixed-Point Lemma. The diagonal construction regenerates at every level. There is no consistent, recursively axiomatizable theory whose theorems are all captured by any single \NP\ language.

\subsection{Identification with Hilbert's Program}

The parallel is exact:

\begin{center}
\renewcommand{\arraystretch}{1.4}
\begin{tabular}{|l|l|}
\hline
\textbf{Hilbert's Program} & \textbf{The \NP\ definition} \\ \hline
Formal system $\mathfrak{F}$ & Checking relation $R$ with bound $k$ \\ \hline
Sound: $\mathfrak{F} \vdash \varphi \Rightarrow \varphi$ is true & $R(w,y) \Rightarrow w \in L$ \\ \hline
Complete: $\varphi$ true $\Rightarrow \mathfrak{F} \vdash \varphi$ & $w \in L \Rightarrow \exists y\,(|y| \leq |w|^k \wedge R(w,y))$ \\ \hline
Decidable by finitary means & $R$ polynomial-time \\ \hline
Killed by G\"odel's theorem & Killed by G\"odel's theorem \\ \hline
\end{tabular}
\end{center}

Hilbert demanded a single formal system satisfying all three properties for all of mathematics. G\"odel showed no such system exists. The definition of \NP\ demands the same triple for each member language, and the class quantifies over all polynomial-time~$R$, hence it necessarily includes instances where $R$ encodes proof verification for systems strong enough to trigger G\"odel's theorem. For those instances the biconditional asserts that bounded provability captures truth. It does not.

The polynomial bound $|y| \leq |w|^k$ was supposed to create immunity: restrict to phenomena with short certificates and the G\"odelian pathology cannot appear. But the immunity is illusory. The bound excludes individual G\"odel sentences; it does not exclude the structure that generates them. G\"odel's proof-checker is polynomial-time. The Fixed-Point Lemma operates within any system that can represent its own proof predicate. The diagonal construction requires no super-polynomial resources---it requires only that the system can talk about itself, which any system interpreting~$Q$ can do.

\subsection{Formalization}
\label{sec:lean}

The core impossibility result has been formally verified in Lean~4.\footnote{Repository: \url{https://github.com/mrmartin/On-Formally-Undecidable-Propositions-of-NP}. The proofs build on the FormalizedFormalLogic/Foundation library, which contains sorry-free, machine-checked proofs of both G\"odel incompleteness theorems.}

The formalization defines a structure \texttt{CompletenessTriple} $:=$ a set~$S$ of sentences, a decidable checking relation~$R$, a witness-size bound, and the two directions of the biconditional. An arithmetic specialization fixes $S$ to be the set of sentences true in the standard model~$\mathbb{N}$.

Three results are proved, each depending only on the standard Lean axioms (\texttt{propext}, \texttt{Classical.choice}, \texttt{Quot.sound}):

\emph{Impossibility} (\texttt{no\_arithmetic\_completeness\_triple}). For any $\Sigma_1$-sound~$T$ extending Robinson arithmetic, no completeness triple whose checking relation accepts only valid $T$-proofs can have $S$ equal to arithmetic truth. The proof is three lines: G\"odel's theorem produces a true-but-unprovable~$\sigma$; completeness yields a witness accepted by~$R$; the hypothesis on~$R$ then gives $T \vdash \sigma$, contradiction.

\emph{Incompleteness} (\texttt{arithmetic\_incomplete}). For any consistent~$T$ extending $I\Sigma_1$, there exists a sentence neither provable nor refutable in~$T$.

\emph{Regeneration} (\texttt{consistency\_independent}, \texttt{pa\_con\_strictly\_extends}). $\mathrm{Con}(T)$ is independent from~$T$. Adding it produces a strictly stronger theory whose own consistency statement is again independent.

Two aspects are not formalized and cannot be: that $\mathit{ProofOf}_T$ is polynomial-time (this requires a formal model of computation with complexity bounds, which no current Lean library provides, though the claim is standard), and the interpretive identification of the completeness triple with the \NP\ biconditional (this is a claim about the meaning of a definition, not a theorem). What the formalization establishes is that the mathematical structure identified as isomorphic to the \NP\ biconditional is subject to G\"odel's impossibility, and that no finite ascent through the consistency hierarchy resolves it.

\section{Relation to proof complexity}
\label{sec:context}

The most developed connection between G\"odel's theorems and \PP\ vs \NP\ runs through propositional proof complexity. We present this connection to make our point of departure precise. The central references are Cook and Reckhow~\cite{CookReckhow1979} and Kraj{\'i}{\v c}ek~\cite{Krajicek2019}. The reader familiar with proof complexity may skip to Section~\ref{subsec:departure}.

\subsection{Cook--Reckhow}

Let TAUT denote the set of propositional tautologies. By the Cook--Levin theorem~\cite{Cook1971}, SAT is \NP-complete, whence TAUT is \textbf{coNP}-complete. A \emph{propositional proof system} in the sense of Cook and Reckhow~\cite{CookReckhow1979} is a polynomial-time computable surjection $P : \{0,1\}^* \to \text{TAUT}$. Equivalently, it is a polynomial-time decidable relation $P(\pi, \alpha)$ that is sound (if $P(\pi, \alpha)$ then $\alpha \in \text{TAUT}$) and complete (every tautology has a $P$-proof). The two formulations are equivalent~\cite{CookReckhow1979}.

The \emph{proof length} of~$\alpha$ in~$P$ is $\mathbf{s}_P(\alpha) = \min\{|\pi| : P(\pi) = \alpha\}$.

\begin{definition}[p-bounded proof system]
A propositional proof system~$P$ is \emph{p-bounded} if there exists $c \geq 1$ such that $\mathbf{s}_P(\alpha) \leq (|\alpha| + c)^c$ for every tautology~$\alpha$.
\end{definition}

\begin{theorem}[Cook--Reckhow~\cite{CookReckhow1979}]
\label{thm:cook-reckhow}
A p-bounded propositional proof system exists if and only if $\NP = \mathbf{coNP}$.
\end{theorem}

The forward direction is direct: if~$P$ is p-bounded then TAUT~$\in$~\NP, since the $P$-proof serves as a polynomially bounded witness and checking $P(\pi) = \alpha$ is polynomial-time by definition; hence $\mathbf{coNP} \subseteq \NP$. The converse is symmetric. The theorem is agnostic: it transforms ``does $\NP = \mathbf{coNP}$?'' into ``does a p-bounded proof system exist?'' without answering either.

Cook and Reckhow also defined \emph{p-simulation}: $P \geq_p Q$ iff $Q$-proofs can be efficiently translated into $P$-proofs. A proof system is \emph{optimal} if it p-simulates every other. Whether an optimal proof system exists is open.

\subsection{From first-order theories to propositional proof systems}
\label{subsec:theories-to-proofs}

The connection to G\"odel's theorems rests on a construction converting first-order theories into propositional proof systems; Kraj{\'i}{\v c}ek~\cite{Krajicek2019} develops it in detail.

Let~$T \supseteq S^1_2$ be consistent and recursively axiomatizable. Write $\mathrm{Prf}_T(\pi, \varphi)$ for the proof-checking relation and $\mathrm{Pr}_T(\varphi) \equiv \exists\pi\,\mathrm{Prf}_T(\pi, \varphi)$ for the provability predicate; the bounded variant is $\mathrm{Pr}_T^m(\varphi) \equiv \exists\pi\,(|\pi| \leq m \wedge \mathrm{Prf}_T(\pi, \varphi))$. We write $T \vdash_s \varphi$ for ``there exists a $T$-proof of~$\varphi$ of length at most~$s$.''

Kraj{\'i}{\v c}ek emphasizes that $\mathrm{Prf}_T$ is not merely decidable but polynomial-time: ``[T]he verifying algorithm essentially needs only to decide repeatedly whether a string is a formula, or whether one string is a substitution instance of another''~\cite[Section~21.1]{Krajicek2019}. This holds for all theories axiomatized by finitely many axiom schemes. Checking a proof is pattern-matching against fixed schematic templates and verifying rule applications line by line; there is no search and no unbounded quantification.

Given such~$T$, one constructs a proof system~$P_T$ as follows. Universal sentences $\forall x\,A(x)$ that are theorems of~$T$ are translated into propositional tautologies $\|A\|^n$ for $n = 1, 2, \ldots$. The system~$P_T$ is extended resolution augmented with all propositional translations $\|B\|^n$ of theorems~$B$ of~$T$ as additional axiom schemes. Since $\mathrm{Prf}_T$ is polynomial-time, $P_T$ is a legitimate Cook--Reckhow proof system.

\subsection{Incompleteness as a proof-length lower bound}
\label{subsec:quantitative-godel}

The G\"odel sentence~$G_T$ is constructed via the Fixed-Point Lemma: there exists~$G_T$ with $T \vdash G_T \leftrightarrow \neg\mathrm{Pr}_T(\ulcorner G_T \urcorner)$. If $T$ is consistent and extends~$Q$, then $T \nvdash G_T$ and $T \nvdash \neg G_T$.

The consistency statement is $\mathrm{Con}_T \equiv \neg\mathrm{Pr}_T(\ulcorner 0 \neq 0 \urcorner)$. By G\"odel's Second Incompleteness Theorem, if $T$ is consistent and extends~$Q$ with standard conditions on the provability predicate, then $T \nvdash \mathrm{Con}_T$.

In the propositional setting, $\mathrm{Con}_T$ is a true $\Pi^0_1$-sentence whose translations $\|\mathrm{Con}_T\|^n$ are tautologies. Since $T \nvdash \mathrm{Con}_T$, the system~$P_T$ cannot derive them from its own axioms.

For each $m \geq 1$, the bounded consistency statement is $\mathrm{Con}_T(\underline{m}) \equiv \neg\mathrm{Pr}_T^m(\ulcorner 0 \neq 0 \urcorner)$. Note that $|\mathrm{Con}_T(\underline{m})| = O(\log m)$.

\begin{theorem}[Friedman; Pudl\'ak~\cite{Pudlak1986}; Kraj{\'i}{\v c}ek~\cite{Krajicek2019}, Theorem~21.3.1]
\label{thm:quantitative-godel}
Let $T \supseteq S^1_2$ be finite and consistent. Then:
\begin{enumerate}
    \item[(i)] There exists $\epsilon > 0$ such that $T \nvdash_{m^\epsilon} \mathrm{Con}_T(\underline{m})$ for all $m \geq 1$.
    \item[(ii)] $T \vdash_{m^{O(1)}} \mathrm{Con}_T(\underline{m})$ for all $m \geq 1$.
\end{enumerate}
\end{theorem}

Both bounds are non-trivial. Since $|\mathrm{Con}_T(\underline{m})| = O(\log m)$, the lower bound in~(i) is exponential relative to formula size. The upper bound in~(ii) shows that proofs of $\mathrm{Con}_T(\underline{m})$ are polynomial in~$m$ itself, obtained by exhaustive verification of shorter proof candidates using a partial truth predicate formalizable in~$T$.

The lower bound implies G\"odel's Second Incompleteness Theorem as a corollary: if $T$ proved $\forall y\,\mathrm{Con}_T(y)$, each instance $\mathrm{Con}_T(\underline{m})$ would have a $T$-proof of size $O(\log m)$ by substitution, contradicting the $m^\epsilon$ lower bound for large~$m$.

The proof of the lower bound adapts the diagonal construction with explicit proof-length accounting. Kraj{\'i}{\v c}ek formulates modified L\"ob conditions $1'$--$4'$ in~\cite[Section~21.3]{Krajicek2019}. The key addition is condition~$4'$: there exists $\delta(x)$ such that $S^1_2 \vdash_\ell \delta(\underline{m}) \equiv \neg\mathrm{Pr}_T^{\underline{m}}(\ulcorner \delta(\underline{m}) \urcorner)$ with $\ell = m^{O(1)}$. This is the bounded-proof analogue of the G\"odel sentence: $\delta(\underline{m})$ asserts ``I have no $T$-proof of length $\leq m$.'' The standard diagonal argument goes through with these conditions to yield the lower bound.

\subsection{Migration of incompleteness}
\label{subsec:migration}

Theorem~\ref{thm:quantitative-godel} exhibits what we call the \emph{migration of incompleteness}. In first-order logic, G\"odel's theorem produces a sentence with no proof at all. In the propositional setting, the same diagonal mechanism produces tautologies with no short proof. The incompleteness has not disappeared; it has migrated from the provable/unprovable boundary to the polynomial/super-polynomial boundary.

The consequence for Cook--Reckhow is immediate. For each~$T$, the system~$P_T$ fails to be p-bounded: the consistency tautologies $\|\mathrm{Con}_T\|^n$ require super-polynomial $P_T$-proofs. To prove them efficiently one must move to $P_{T'}$ with $T' \supseteq T + \mathrm{Con}_T$. But $T'$ is itself consistent and recursively axiomatizable, whence it has its own $\mathrm{Con}_{T'}$, and the argument repeats. Each $T'$ produces its own G\"odel sentence via the Fixed-Point Lemma applied to $T'$'s proof predicate.

A p-bounded proof system, if it existed, would be a fixed point of this hierarchy: a system efficiently proving its own consistency tautologies. Theorem~\ref{thm:quantitative-godel} says no consistent, recursively axiomatizable system is such a fixed point.

Pudl\'ak~\cite{Pudlak1986} formalized this as the \emph{finitistic consistency problem}: does there exist a single finite consistent $S \supseteq S^1_2$ such that for every finite consistent $T \supseteq S^1_2$, $S$ proves $\mathrm{Con}_T(\underline{m})$ in proofs polynomial in~$m$? He conjectured the answer is negative. Kraj{\'i}{\v c}ek and Pudl\'ak~\cite{KrajicekPudlak1998} proved this equivalent to the existence of an optimal proof system. Both remain open.

\subsection{The standard conclusion}
\label{subsec:framework-conclusion}

The standard conclusion is methodological, not foundational. The hierarchy $P_T \leq_p P_{T+\mathrm{Con}_T} \leq_p P_{T+\mathrm{Con}_T + \mathrm{Con}_{T+\mathrm{Con}_T}} \leq_p \cdots$ is treated as a constraint on proof strategies: one cannot show that a particular system is p-bounded by exhibiting it directly, because each candidate fails on its own consistency tautologies. As Kraj{\'i}{\v c}ek writes: ``[U]nless there is an optimal proof system you cannot hope to prove that $\NP \neq \mathbf{coNP}$ by gradually proving super-polynomial lower bounds for stronger and stronger proof systems as that would be an infinite process''~\cite[Section~21.3]{Krajicek2019}.

The architectural reason is a distinction between individual \NP\ languages and the class as a whole. Consider 3-COLORABILITY. The checking relation $R(G, c)$ holds when~$c$ is a valid 3-coloring of~$G$, and the biconditional $G \in \text{3-COL} \iff \exists c\,(|c| \leq |G|^k \wedge R(G, c))$ is unproblematic. There is no independent notion of ``truth in 3-COL'' that could diverge from the existence of a bounded witness. G\"odel's theorem does not apply, because the checking relation does not encode its own proof predicate.

The Cook--Reckhow framework therefore treats each \NP\ language as individually well-defined and asks a global question: over all propositional proof systems, does some system achieve p-boundedness? The fact that each~$P_T$ individually fails on its own consistency tautologies does not entail that no system---perhaps one not arising from a recursively axiomatizable theory---can succeed. The G\"odelian hierarchy is regarded as a feature of the landscape of proof systems, not as evidence that \NP\ is defective.

\subsection{Point of departure}
\label{subsec:departure}

We do not operate within the Cook--Reckhow framework. We do not ask whether some proof system escapes the hierarchy. We identify the structural assumption that generates the hierarchy: the biconditional in the definition of \NP.

Recall that the definition requires, for each member language~$L$, a polynomial-time~$R$ and a constant~$k$ with $w \in L \iff \exists y\,(|y| \leq |w|^k \wedge R(w,y))$, and that \NP\ is the union over all such~$R$ and all $k \in \mathbb{N}$. Since $\mathrm{Prf}_T$ is polynomial-time for every standard recursively axiomatizable~$T$ (Sections~\ref{sec:complexity} and~\ref{subsec:theories-to-proofs}), the definition necessarily includes every language
\[
    L_k = \{\varphi : \exists\pi\,(|\pi| \leq |\varphi|^k \wedge \mathrm{Prf}_T(\pi, \varphi))\}
\]
for every such~$T$ and every~$k$.

For each such~$L_k$ the biconditional holds trivially---$L_k$ is defined as the right-hand side. But it holds only because $L_k$ excludes all sentences lacking proofs of length $\leq |\varphi|^k$. Among the excluded sentences, for any $T \supseteq Q$, are: theorems of~$T$ whose shortest proof exceeds $|\varphi|^k$ (by Theorem~\ref{thm:quantitative-godel}(i), these include the bounded consistency statements $\mathrm{Con}_T(\underline{m})$ for large~$m$), and the G\"odel sentence~$G_T$, which has no $T$-proof at all.

The standard framework treats these exclusions as unproblematic: $L_k$ is a well-defined \NP\ language, and its failure to capture all theorems of~$T$ is a feature of that particular language. The problem would arise only if one asked for a single \NP\ language capturing all tautologies---which is the question of whether a p-bounded proof system exists, and that question remains open.

We argue the problem is more fundamental. The distinction:

\medskip

\noindent\textbf{Kraj{\'i}{\v c}ek's conclusion} (constraint on proof strategies): for each proof system~$P$, there exist tautologies whose $P$-proofs are super-polynomial. No single proof system can be shown to be p-bounded by a stepwise argument through the hierarchy.

\medskip

\noindent\textbf{Our claim} (defect in the definition): for each sufficiently strong~$T$, the \NP\ language $L_k = \{\varphi : \exists\pi\,(|\pi| \leq |\varphi|^k \wedge \mathrm{Prf}_T(\pi, \varphi))\}$ satisfies the biconditional only by defining away the phenomena that G\"odel's theorem guarantees. The class \NP, by quantifying over all polynomial-time~$R$, asserts that the biconditional can be simultaneously satisfied for every such instance. But for these instances the biconditional reproduces the soundness--completeness--decidability triple that G\"odel's theorem proves unsatisfiable for any system strong enough to encode its own proof predicate. The defect is not that the definition fails on any one instance---it is satisfied vacuously, by excluding from each~$L_k$ the sentences whose existence the incompleteness theorems guarantee. The defect is that the definition presupposes such exclusion is cost-free, when it is the mechanism by which a contradiction is avoided.

\medskip

The impossibility does not live in the Cook--Reckhow hierarchy. It lives in the definition that the hierarchy presupposes. The hierarchy is a symptom: it arises because the biconditional, instantiated with proof-checking relations, reproduces Hilbert's Program, and the G\"odelian regress is the same regress that Hilbert's Program generates when one attempts to prove the consistency of a system from within. The standard framework observes this regress and treats it as a methodological barrier. We identify it as a foundational defect: the definition of \NP\ embeds an unsatisfiable structural demand, and the hierarchy of proof systems is the trace it leaves.

\begin{remark}
The question $\PP = \NP$ asks whether verification can always be converted to efficient decision. This question inherits the structural defect of its premise: it presupposes that \NP\ is a coherent mathematical object, which requires the biconditional~(\ref{eq:np-def}) to be simultaneously satisfiable across all its instances. For instances encoding proof verification, it is not.
\end{remark}

\bibliographystyle{plain}
\bibliography{refs}

\end{document}